\documentclass[submitting]{nst}

\usepackage{subfigure,dcolumn}
\usepackage{epstopdf}
\usepackage{mhchem}
\usepackage{upgreek}

\let\linenumbers\relax

\begin{document}

\title{Response of GAGG(Ce) scintillators to different charged particles}

\author{Kai-Jie Zhou}
\affiliation{School of Physics and State Key Laboratory of Nuclear Physics and Technology, Peking University, Beijing 100871, China}

\author{Zai-Hong Yang}
\email[Corresponding author, ]{zaihong.yang@pku.edu.cn}
\affiliation{School of Physics and State Key Laboratory of Nuclear Physics and Technology, Peking University, Beijing 100871, China}

\author{Hooi-Jin Ong}
\affiliation{State Key Laboratory of Heavy Ion Science and Technology, Institute of Modern Physics, Chinese Academy of Sciences, Lanzhou 730000, China}
\affiliation{School of Nuclear Science and Technology, University of Chinese Academy of Sciences, Beijing 100049, China}
\affiliation{Joint Department for Nuclear Physics, Lanzhou University and Institute of Modern Physics, Chinese Academy of Sciences, Lanzhou 730000, China}
\affiliation{Research Center for Nuclear Physics, Osaka University, Ibaraki, Osaka 567-0047, Japan}
\affiliation{Nishina Center for Accelerator-Based Science, RIKEN, 2-1 Hirosawa, Wako, 351-0198 Saitama, Japan}

\author{Qi-Te Li}
\affiliation{School of Physics and State Key Laboratory of Nuclear Physics and Technology, Peking University, Beijing 100871, China}

\author{Ze-Yu Du}
\affiliation{School of Physics and State Key Laboratory of Nuclear Physics and Technology, Peking University, Beijing 100871, China}

\author{Jia-Wei Bian}
\affiliation{School of Physics and State Key Laboratory of Nuclear Physics and Technology, Peking University, Beijing 100871, China}

\author{Cheng Wang}
\affiliation{School of Physics and State Key Laboratory of Nuclear Physics and Technology, Peking University, Beijing 100871, China}

\author{Jin-Ning Li}
\affiliation{School of Physics and State Key Laboratory of Nuclear Physics and Technology, Peking University, Beijing 100871, China}

\author{Satoru Terashima}
\affiliation{State Key Laboratory of Heavy Ion Science and Technology, Institute of Modern Physics, Chinese Academy of Sciences, Lanzhou 730000, China}
\affiliation{Research Center for Nuclear Physics, Osaka University, Ibaraki, Osaka 567-0047, Japan}
\affiliation{Nishina Center for Accelerator-Based Science, RIKEN, 2-1 Hirosawa, Wako, 351-0198 Saitama, Japan}

\author{Bing-Feng Lv}
\affiliation{State Key Laboratory of Heavy Ion Science and Technology, Institute of Modern Physics, Chinese Academy of Sciences, Lanzhou 730000, China}
\affiliation{School of Nuclear Science and Technology, University of Chinese Academy of Sciences, Beijing 100049, China}

\author{Xuan Wang}
\affiliation{School of Physics, Beihang University, Beijing 100191, China}

\author{Zhi-Chao Zhang}
\affiliation{State Key Laboratory of Heavy Ion Science and Technology, Institute of Modern Physics, Chinese Academy of Sciences, Lanzhou 730000, China}

\author{Hong-Na Liu}
\affiliation{School of Physics and Astronomy, Beijing Normal University, Beijing 100875, China}

\author{Yu-Yang Yu}
\affiliation{School of Physics and Astronomy, Beijing Normal University, Beijing 100875, China}

\author{Peng-Jie Li}
\affiliation{Key Laboratory of Nuclear Physics and Ion-beam Application (MOE), Institute of Modern Physics, Fudan University, Shanghai 200433, China}

\author{Ye-Lei Sun}
\affiliation{School of Physics, Beihang University, Beijing 100191, China}

\begin{abstract}

Charged-particle telescopes play an essential role in studies of nuclear clustering, which require high-resolution detection of charged breakup fragments. GAGG(Ce) is a promising scintillator for charged-particle detection at high counting rates because of its high light yield, fast response, and non-hygroscopic nature. However, systematic data on its light-output response to different charged particles remain scarce. In this work, we present a systematic study of the scintillation light-output response of GAGG(Ce) crystals to various charged particles and quantify the associated quenching behavior. The relationship between the scintillation light output and the deposited energy is established over a broad range of particle species and energies, providing essential experimental input for response modeling and energy calibration of GAGG(Ce)-based charged-particle detector systems.
\end{abstract}


\maketitle

\section{Introduction}\label{sec.I}

The study of exotic nuclear structures is a central topic in nuclear physics~\cite{bib:1}. With the continuous development of radioactive ion beam (RIB) facilities~\cite{haif,hirfl,riken,fair,spiral,frib}, nuclear structure studies have expanded toward regions far from the stability line, where nuclei often behave as weakly bound open quantum systems. The structure of such nuclei can deviate significantly from traditional nuclear structure model expectations, giving rise to exotic structure phenomena such as neutron halos and clustering~\cite{bib:2,bib:3,bib:4}. Nuclear clustering, where nucleons form correlated substructures such as $\alpha$ clusters, is a key emergent feature of nuclear many-body systems~\cite{bib:1,bib:4}. Well-known examples include the Hoyle state of \ce{^{12}C} with a 3-$\alpha$ cluster structure~\cite{bib:5} and molecular-like cluster structures in Be and C isotopes~\cite{bib:6,bib:a}. Studies of nuclear clustering can provide important constraints on nuclear interactions and theoretical descriptions of nuclei, as well as the properties of nuclear matter~\cite{bib:2,bib:8,bib:b}.

Experimental studies of clustering are typically carried out using inelastic scattering and knockout reactions, which require precise measurements of energies and emission angles of charged-particle reaction products~\cite{bib:3}. Consequently, charged-particle telescope arrays with high energy and angular resolution have become indispensable tools for such investigations. A variety of charged-particle telescope systems based on combinations of silicon detectors and scintillators have been developed worldwide, including LASSA~\cite{bib:9}, MUST2~\cite{bib:10}, HiRA~\cite{bib:11}, FARCOS~\cite{bib:12,bib:13}, and TOGAXSI~\cite{bib:14}. 

Motivated by these experimental requirements, we have developed a charged-particle telescope array based on double-sided silicon detectors (DSSD) coupled with Ce-doped \ce{Gd_3Al_2Ga_3O_{12}} scintillators, GAGG(Ce). The system is designed to provide particle identification and energy measurement over a broad dynamic range, and is optimized for multi-particle coincidence measurements relevant to studies of nuclear cluster structures.

GAGG(Ce) is a relatively recent inorganic scintillator material featuring high density ($\sim$6.63 \si{\gram\per\cubic\centi\metre}), high light yield ($\sim$50,000 photons/MeV), fast decay time ($\sim$60        --100 ns), and excellent chemical and mechanical stability~\cite{bib:15}. Unlike conventional NaI(Tl) and CsI(Tl) scintillators, GAGG(Ce) is non-hygroscopic. Previous studies reported dominant decay time constants of approximately 60 - 100 ns~\cite{bib:16}, making GAGG(Ce) particularly well suited for long-term operation and high count-rate experimental environments.


Despite these favorable properties for nuclear physics experiments, systematic studies of the scintillation response of GAGG(Ce) scintillators remain limited. Existing measurements have primarily focused on low-energy $\alpha$ particles and protons~\cite{bib:17,bib:18}, while systematic investigations covering higher energies and heavier ions are scarce. Such information is essential for applications involving energetic heavy charged particles, such as quasi-free knockout and inelastic scattering reactions with RIBs.

In this work, we present a systematic study of the light-output response of GAGG(Ce) scintillators to various charged particles, including \ce{^4 He}, \ce{^{10}Be}, \ce{^{10}C}, \ce{^{11}C}, \ce{^{16}C}, and \ce{^{18}O} with energies above 100 MeV. The scintillation efficiency, expressed as $dL/dE$, is analyzed as a function of the linear energy transfer (LET), $dE/dx$. These results provide essential experimental input for understanding quenching effects in GAGG(Ce) scintillators and support for their application in charged-particle detector systems for nuclear structure studies.

\section{Experimental setup and detector system}\label{sec.II}
The experiment was carried out at the RIBLL1 beam line~\cite{bib:19} of the Institute of Modern Physics, Chinese Academy of Sciences. Secondary beams of \ce{^{4}He}, \ce{^{10}Be}, \ce{^{10}C}, and \ce{^{11}C} were produced via fragmentation of a \ce{^{12}C} primary beam with an energy of approximately 80 MeV/nucleon, while the \ce{^{16}C} secondary beam was produced using an \ce{^{18}O} primary beam with an energy of approximately 60 MeV/nucleon. All secondary beams were purified and transported to the experimental terminal by the RIBLL1 beam line. An energy-degraded \ce{^{18}O} primary beam was also delivered directly to the GAGG(Ce) crystals for dedicated response measurements. The beam intensity was limited to approximately $1.5 \times 10^5$ pps in order to reduce the dead time of the data acquisition system.

\begin{figure}[h]
	\includegraphics[width=1\hsize]{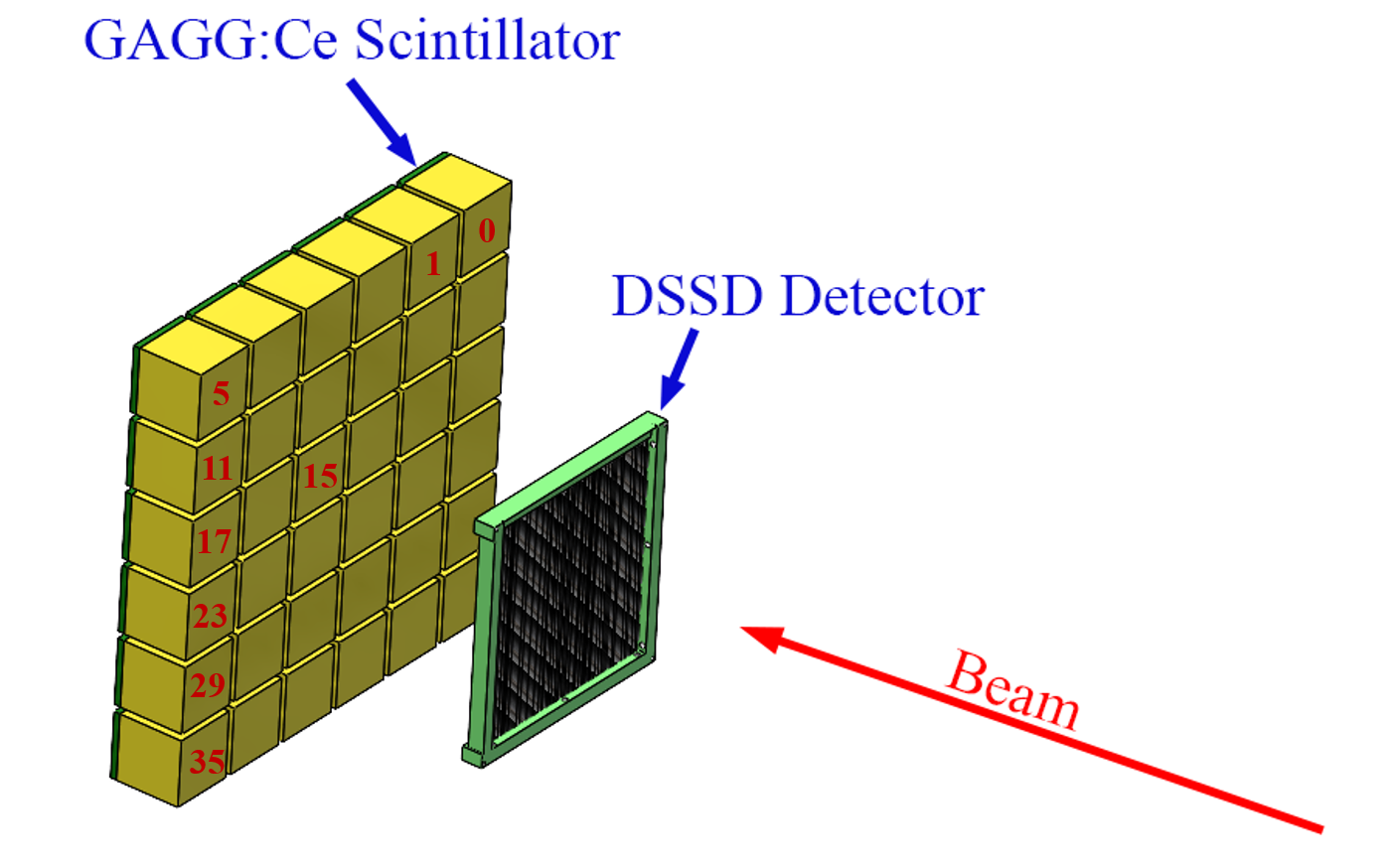}
	\caption{Schematic view of the experimental setup at RIBLL1 beamline}
	\label{fig1}
\end{figure}

A schematic view of the experimental setup is shown in Fig.~\ref{fig1}. An $\Delta E-E$ telescope was constructed using a DSSD as the $\Delta E$ detector, followed by an array of GAGG(Ce) scintillators serving as the E detector. The $\Delta E$ detector was a \SI{500}{}-\si{\micro\metre}-thick DSSD with an active area of $\SI{64}{\milli\metre} \times \SI{64}{\milli\metre}$. The E detector consisted of an array of 36 GAGG(Ce) scintillator crystals, labeled from 0 to 35 as indicated in Fig.~\ref{fig1}. All scintillators were manufactured by Scionix (Netherlands), with individual crystal dimensions of $\SI{20}{\milli\metre} \times \SI{20}{\milli\metre} \times \SI{25}{\milli\metre}$. The crystal surfaces were polished and exhibited good optical transparency upon visual inspection. All analyzed crystals were from the same production batch.

Beams of different particle species and energies were directly injected into the telescope. After passing through the DSSD, the residual energies of the incident particles were fully deposited in the GAGG(Ce) scintillators. The particle species used in the measurement and the corresponding deposited energies in the GAGG(Ce) scintillators and the energy losses in the DSSD, calculated using LISE++ while accounting for the energy loss in the secondary target, are listed in Table~\ref{tab:1}. The associated uncertainties mainly arise from the beam-energy spread, DSSD thickness tolerance, and material thicknesses along the particle path. The combined uncertainty is included in Table~\ref{tab:1}.

\begin{table*}[!htb]
	\centering
	\caption{Particle species and deposited energies in GAGG calculated using LISE++}
	\label{tab:1}
	\renewcommand{\arraystretch}{1.3}
	\begin{tabular*}{\textwidth}{@{\extracolsep{\fill}} lcccccccc}
		\toprule
		Particle & \ce{^4 He} & \ce{^4 He} & \ce{^{10}Be} & \ce{^{10}C} & \ce{^{11}C} & \ce{^{16}C} & \ce{^{16}C} & \ce{^{18}O} \\
		\midrule
		deposited energy in GAGG(Ce) (MeV) & 116.5(6) & 212.9(11) & 619(3) & 539(3) & 622(3) & 846(4) & 271.0(14) & 161.1(8) \\
		\midrule
		deposited energy in DSSD (MeV) & 6.8(2) & 4.3(1) &  15.4(1) & 37.5 & 34.0(2) & 38.6(2) & 86.8(9) & 196(4) \\
		\bottomrule
	\end{tabular*}
\end{table*}

\begin{figure}[!htb]
	\subfigure{
		\label{fig2a}
		\includegraphics[width=0.45\linewidth]{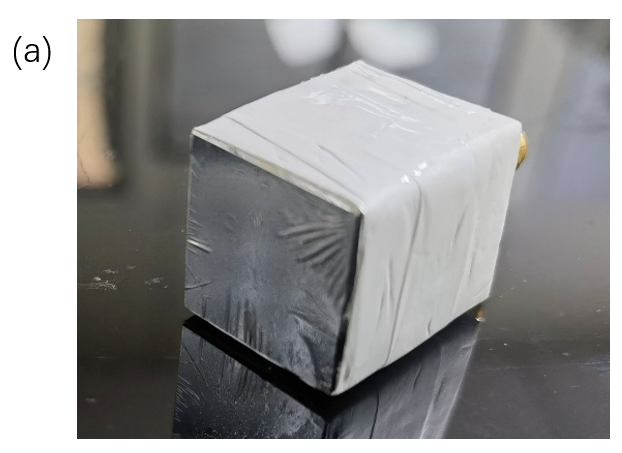}
	}
	\subfigure{
		\label{fig2b}
		\includegraphics[width=0.45\linewidth]{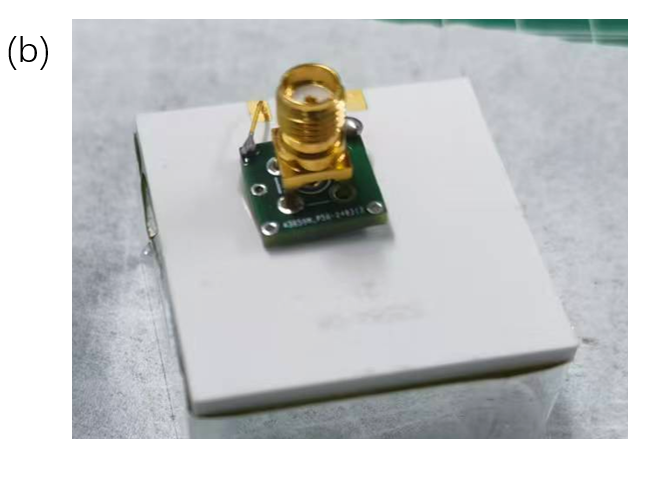}
	}
	\caption{Photograph of (a) a fully assembled scintillator unit (b) and the photodiode}
	\label{figgagg}
\end{figure}

To enhance light collection efficiency, the lateral surfaces of each GAGG(Ce) crystal were wrapped with enhanced specular reflector (ESR) film. The beam-incident surface was covered with a \SI{2}{}-\si{\micro\metre}-thick aluminized Mylar foil, which provided effective light shielding while minimizing energy loss of the incident particles. The rear surface of each scintillator was coupled to a Hamamatsu S3590-08 photodiode (PD), with an active area of $\SI{10}{\milli\metre} \times \SI{10}{\milli\metre}$, using optical silicone grease. Since the PD active area ($\SI{10}{\milli\metre} \times \SI{10}{\milli\metre}$) is smaller than the crystal face ($\SI{20}{\milli\metre} \times \SI{20}{\milli\metre}$), the PD was coupled to the central region of the crystal and the uncovered region on the coupling surface was also wrapped with ESR film to improve light collection, thereby reducing the position dependence of light collection. A photograph of a fully assembled scintillator unit is shown in Fig.~\ref{fig2a}, and a photograph of the photodiode is shown in Fig.~\ref{fig2b}.

The output signals from the photodiodes were amplified using charge-sensitive preamplifiers (SPA-02)~\cite{bib:20} before digitization. The preamplifiers feature a compact design and high channel density, which allows convenient installation inside the chamber, and significantly simplifies cabling and experimental setup. The decay time constant of the preamplifiers is set to \SI{20}{\micro\second}, optimized for the requirements of the present measurements.

Signal digitization and data acquisition were performed using 16-channel Pixie-16 digital pulse processors manufactured by XIA LLC. The Pixie-16 modules were installed in a Compact PCI/PXI-based chassis and are capable of transferring data from the module memory to the host computer at rates up to \SI{109}{\mega\byte\per\second}~\cite{bib:21}. Digital trapezoidal filtering, implemented in the Pixie-16 module, was applied to process the exponentially decaying signals from the charge-sensitive preamplifiers, enabling accurate pulse-height (energy) extraction and timing measurements.

\begin{figure*}[!htb]
	\includegraphics[width=0.9\hsize]{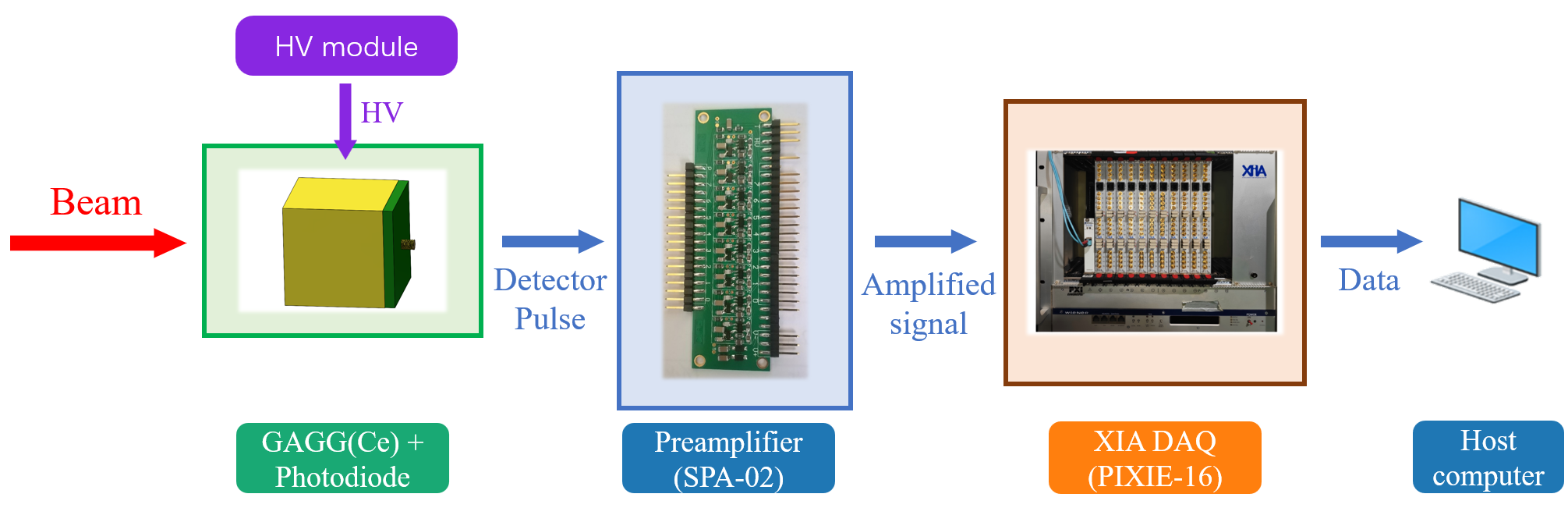}
	\caption{Schematic of the electronics and data-acquisition chain.}
	\label{figdaq}
\end{figure*}

A schematic diagram of the signal chain is shown in Fig.~\ref{figdaq}. Scintillation light produced in each GAGG(Ce) crystal is detected by a PD, and converted into a charge signal, which is subsequently fed into a charge-sensitive preamplifier (SPA-02) for amplification and shaping. The preamplifier outputs are then delivered to the Pixie-16 modules for digitization. The pulse-height and timing information are extracted on an event-by-event basis using digital signal processing, and the processed data are transferred to the host computer for storage and offline analysis.

\begin{figure*}[!htb]
	\subfigure{
		\label{fig4a}
		\includegraphics[width=0.45\linewidth]{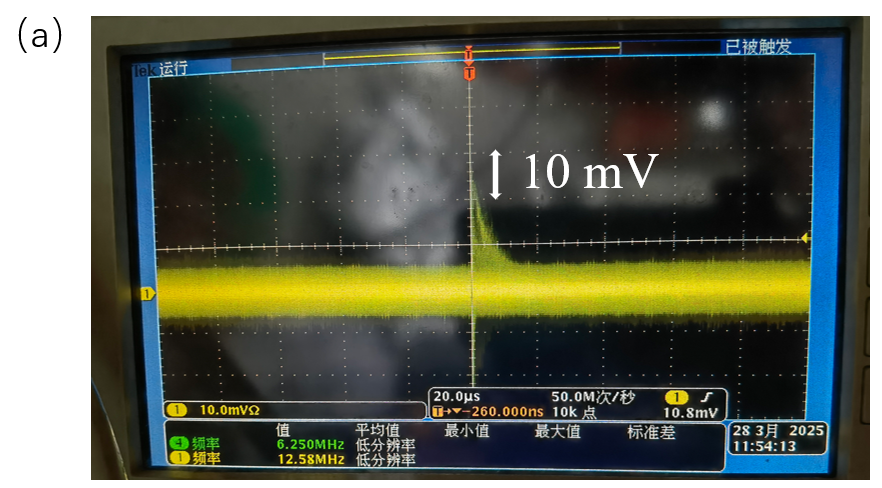}
	}
	\subfigure{
		\label{fig4b}
		\includegraphics[width=0.45\linewidth]{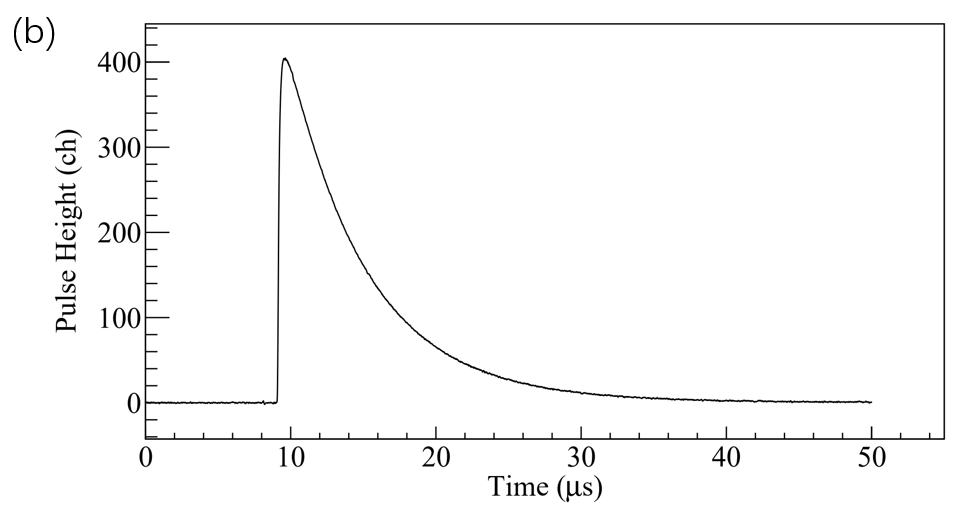}
	}
	\caption{(a) Representative waveform at the output of the charge-sensitive preamplifier for the GAGG(Ce) scintillator, recorded during the cosmic-ray test.
		(b) Average preamplifier-output waveform obtained by peak-aligned averaging of multiple waveforms.}
	\label{figpulse}
\end{figure*}

\section{Pulse shape}\label{sec.III}

Prior to the beam-test experiment, cosmic-ray events were used to examine the signal characteristics and readout performance of the GAGG(Ce) scintillators. A representative waveform recorded with an oscilloscope at the output of the preamplifier is shown in Fig.~\ref{fig4a}. To obtain a typical pulse shape, a large number of waveforms were time-aligned at their peak positions and averaged point by point, as shown in Fig.~\ref{fig4b}. The observed signal pulse width is approximately \SI{20}{\micro\second}, and the baseline fluctuation is within $\pm\SI{5}{\milli\volt}$, indicating stable signal output and acceptable electronic noise levels.  The shaping time is much longer than the intrinsic scintillation decay time of GAGG(Ce), which is within approximately \SI{1}{\micro\second}. Therefore, the recorded waveform tail ($\sim$\SI{20}{\micro\second}) is dominated by the electronics response rather than the intrinsic scintillation decay. Although particle-dependent pulse-shape variations may exist, the present readout configuration is expected to integrate essentially the full scintillation signal.

To accommodate the high count-rate conditions anticipated in RIB experiments, the decay time constant of the charge-sensitive preamplifier was optimized by adjusting the input resistance. As shown in Fig.~\ref{fig4b}, the resulting preamplifier-output pulse width of about \SI{20}{\micro\second} is sufficiently short to handle event rates of $\sim\SI{100}{\kilo\hertz}$, thereby satisfying the experimental requirements.

\section{Relative light output for different particles}\label{sec.IV}

The output signals from the charge-sensitive preamplifiers were digitized and processed by the XIA acquisition system. Each waveform is shaped by a digital trapezoidal filter, and the trapezoid height is extracted as the pulse amplitude, hereafter denoted as the ADC value.

Using crystal No.~15 as a representative example, Fig.~\ref{figbirks} shows the ADC as a function of the deposited energy, where the symbols represent the measured data for all beam settings. A clear quenching behavior is observed: for a given deposited energy, the measured light output decreases as the incident ion becomes heavier, due to the increase of the LET (dE/dx).

For a fixed electronic configuration and shaping parameter setting, the ADC value reflects the integrated charge of the scintillation signal. Although scintillation pulse shapes may vary with particle type, the long shaping time used in this work minimizes sensitivity to such variations. Therefore, the ADC value can be regarded as approximately proportional to the total scintillation light collected by the PD. 

To quantify the quenching effect, we investigated the dependence of the relative light output on particle species and deposited energy, and modeled the scintillation efficiency $dL/dE$ as a function of LET.

\subsection{Birks Model}

As a commonly used empirical description of scintillation quenching, the Birks formula relates the light output per unit path length $dL/dx$ to the stopping power of the charged particle~\cite{bib:22}
\begin{equation}\label{eq1}
	\frac{dL}{dx}=\frac{S\left(dE/dx\right)}{1+kB\left(dE/dx\right)}
\end{equation}
where the parameter $S$ is a scintillation normalization constant and $kB$  is the Birks factor describing the reduction of scintillation efficiency with increasing $dE/dx$. The Birks formula has been successfully used to describe the behavior of various scintillator materials, such as CsI~\cite{bib:23,bib:24}. Dividing Eq.~\ref{eq1} by $dE/dx$ yields
\begin{equation}\label{eq2}
	\frac{dL}{dE}=\frac{a_0}{1+a_1\left(dE/dx\right)}
\end{equation}
where \(a_0\) and \(a_1\) are fitting parameters corresponding to \(S\) and \(kB\), respectively. Here, \(a_0\) gives the overall scale of \(dL/dE\), while \(a_1\) characterizes the quenching strength.

The total light output for a particle depositing an energy $E_0$ in the crystal is obtained by integrating $dL/dE$ over the deposited energy
\begin{equation}\label{eq3}
	L_0=\int_{0}^{E_0}{\frac{a_0}{1+a_1\left(dE/dx\right)}dE^{\prime}}+C
\end{equation}
where $C$ accounts for the overall offset in the relative light-output scale.

\begin{figure}[!htb]
	\includegraphics[width=\hsize]{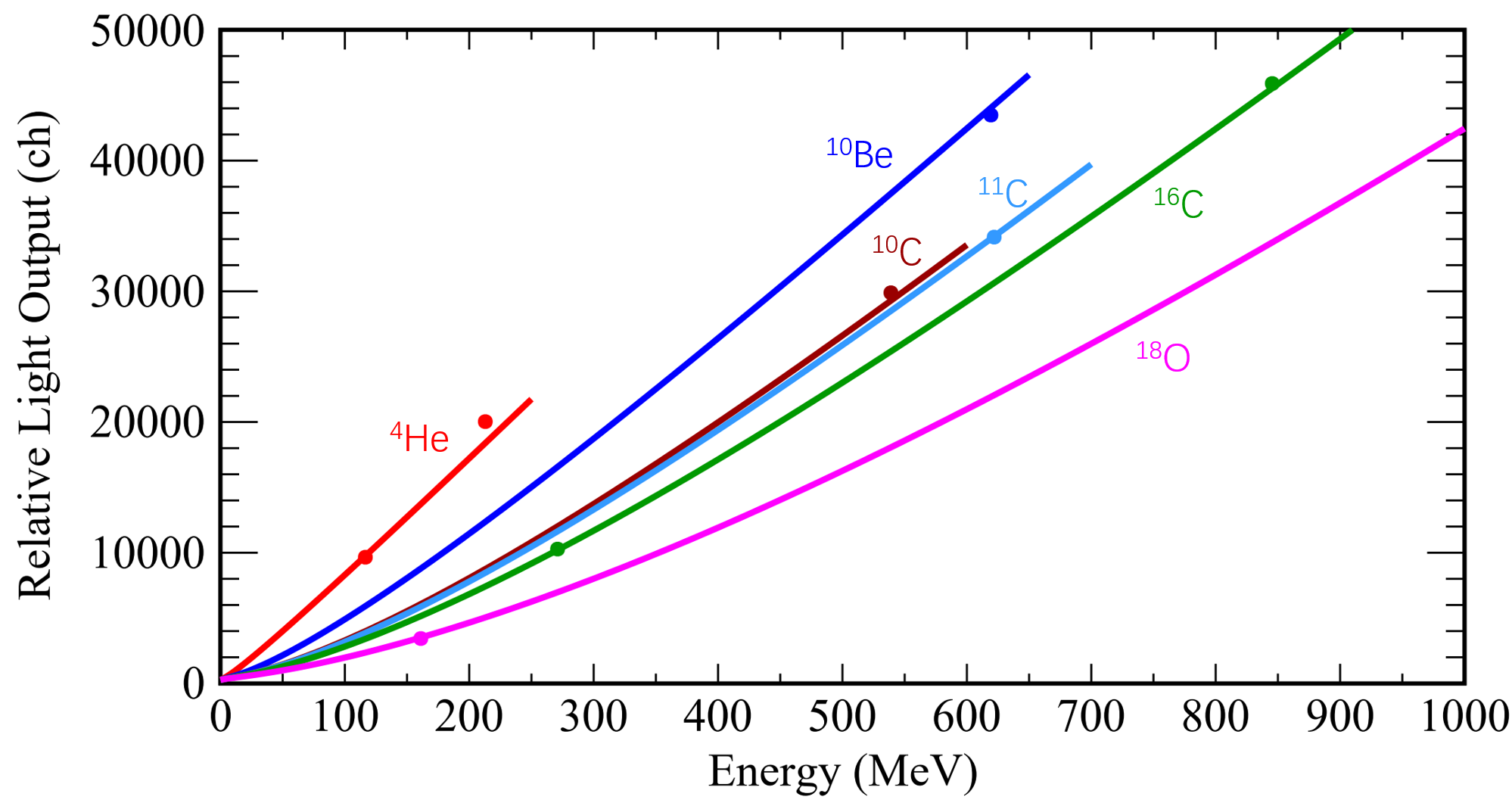}
	\caption{Relative light output as a function of deposited energy for different ions measured with GAGG(Ce) crystal No.~15. Points denote the experimental data, and solid lines show the fit results using the Birks formula~\cite{bib:22}.}
	\label{figbirks}
\end{figure}

For each ion species, $dE/dx$ was obtained from the corresponding stopping-power calculation using LISE++, assuming a GAGG(Ce) density of \SI{6.63}{\gram\per\cubic\centi\metre} and Eq.~\ref{eq3} was used to fit the measured relative light output. Fig.~\ref{figbirks} summarizes the resulting light-output trends for crystal No.~15, and Table~\ref{tab:fitpar15} lists the obtained parameters.Using the fitted Birks parameterization, the deposited energies reconstructed from the measured light outputs differ from the LISE++ calculated values by approximately 1--15\% for the data points used in the fit. This comparison is intended to illustrate the limited descriptive capability of the standard Birks model over the present data set, rather than to represent an independent prediction uncertainty. For a more direct quantitative comparison, the residuals between the fitted and measured relative light outputs are listed in Table~\ref{tab:residuals}.

\subsection{Modified Birks Model}

To improve the description of scintillation response at low LET, Koba \textit{et~al.} introduced an additional term in the denominator, leading to the modified Birks parameterization~\cite{bib:25}:
\begin{equation}\label{eq4}
	\frac{dL}{dE}=\frac{a_0}{1+a_1\left(dE/dx\right)+a_{-1}\left(dE/dx\right)^{-1}}
\end{equation}
Here, \(a_{-1}\) is an empirical correction parameter for the small-\(dE/dx\) region, corresponding in the present work to the relatively high-energy region.

Accordingly, the total light output is given by
\begin{equation}\label{eq5}
	L_0=\int_{0}^{E_0}{\frac{a_0}{1+a_1\left(dE/dx\right)+a_{-1}\left(dE/dx\right)^{-1}}dE^{\prime}}+C
\end{equation}

\begin{figure}[!htb]
	\includegraphics[width=\hsize]{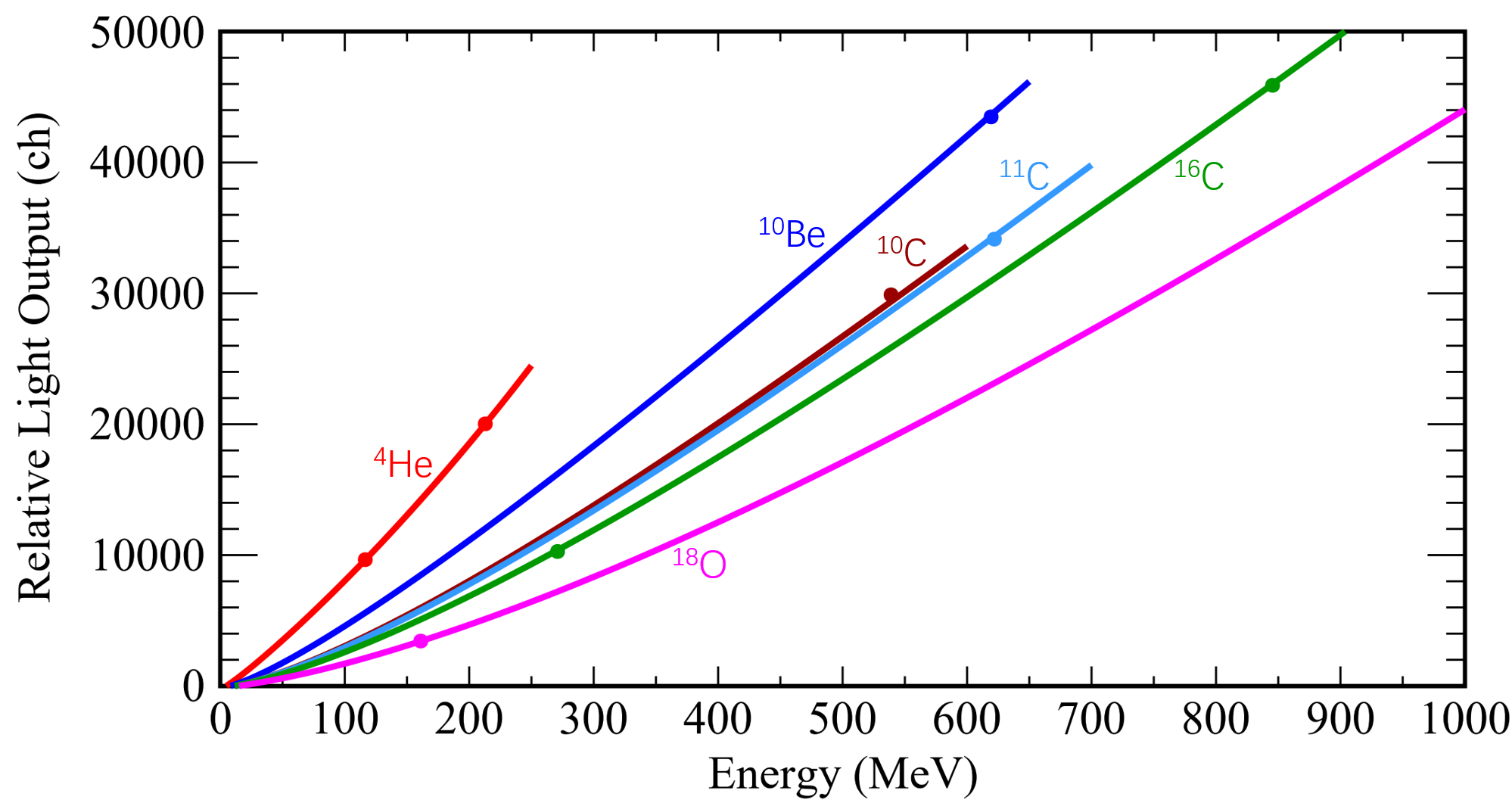}
	\caption{Relative light output as a function of deposited energy for different ions measured with GAGG(Ce) crystal No.~15. Points denote the experimental data, and solid lines show the fit results using the modified Birks parameterization~\cite{bib:25}.}
	\label{figmodified}
\end{figure}

The fit result obtained using the modified Birks model for crystal No.~15 is shown in Fig.~\ref{figmodified}, and the fitted parameters are summarized in Table~\ref{tab:fitpar15}. Compared with the standard Birks model, the modified Birks parameterization provides a visibly improved description of the data. The deviation between reconstructed and expected deposited energies is reduced to within $\SI{3}{\percent}$ over the investigated energy and particle range. For comparison with the standard Birks model, the residuals between the fitted and measured relative light outputs obtained with the modified Birks model are also listed in Table~\ref{tab:residuals}.

Similar improvements are also observed for other crystals in the array, indicating that the modified Birks model provides an adequate empirical description of the scintillation response, particularly the quenching behavior, of GAGG(Ce) scintillators for energetic charged particles. It should be noted that the modified Birks parameterization remains empirical and its applicability outside the measured energy range and to heavier nuclear species requires further validation.

\begin{table}[!htb]
	\caption{Fit parameters for crystal No.~15.}
	\label{tab:fitpar15}
	\begin{tabular*}{\columnwidth}{@{\extracolsep{\fill}} ccc}
		\toprule
		& \multicolumn{2}{c}{parameters} \\
		\cmidrule(l){2-3}
		Model & Parameter & Value \\
		\midrule
		
		& $a_{0}$   & 98.1(4) \\
		Birks & $a_{1}$   & 1.37(2) \\
		& C & 2.5(3)$\times10^2$ \\
		
		\midrule
		
		& $a_{0}$   & 84.9(9) \\
		Modified Birks & $a_{1}$   & 1.00(3) \\
		& $a_{-1}$  & -0.0090(6) \\
		& C & 2.3(5)$\times10^2$\\
		\bottomrule
	\end{tabular*}
\end{table}

\begin{table}[!htb]
	\caption{Comparison of fitting residuals (in ADC channel) for different particles and deposited energies.}
	\label{tab:residuals}
	\begin{tabular*}{\columnwidth}{@{\extracolsep{\fill}} cccc}
		\toprule
		Particle & $E_{\mathrm{dep}}$ (MeV) & \shortstack{Birks model\\residual (ch)} & \shortstack{Modified Birks model\\residual (ch)} \\
		\midrule
		$^{4}$He   & 116.5 & 99.7  & -17.7 \\
		$^{4}$He   & 212.9 & -1643.7 & 7.8 \\
		$^{10}$Be  & 619 & 556.9  & 154.3 \\
		$^{10}$C   & 539 & -601.4 & -545.4 \\
		$^{11}$C   & 622 & 62.5  & 161.5 \\
		$^{16}$C   & 271.0 & -70.3 & 87.2 \\
		$^{16}$C   & 846 & -361.7 & 31.1 \\
		$^{18}$O   & 161.1 & 80.2  & -6.7 \\
		\bottomrule
	\end{tabular*}
\end{table}

\subsection{Crystal-to-crystal comparison}

To evaluate the reproducibility of the scintillation response, seven GAGG(Ce) crystals located near the center of the array were selected and analyzed using the same procedure. Within the modified Birks framework, the parameter $a_1$ provides a convenient quantitative characterization of the quenching strength.

\begin{figure}[!htb]
	\includegraphics[width=\hsize]{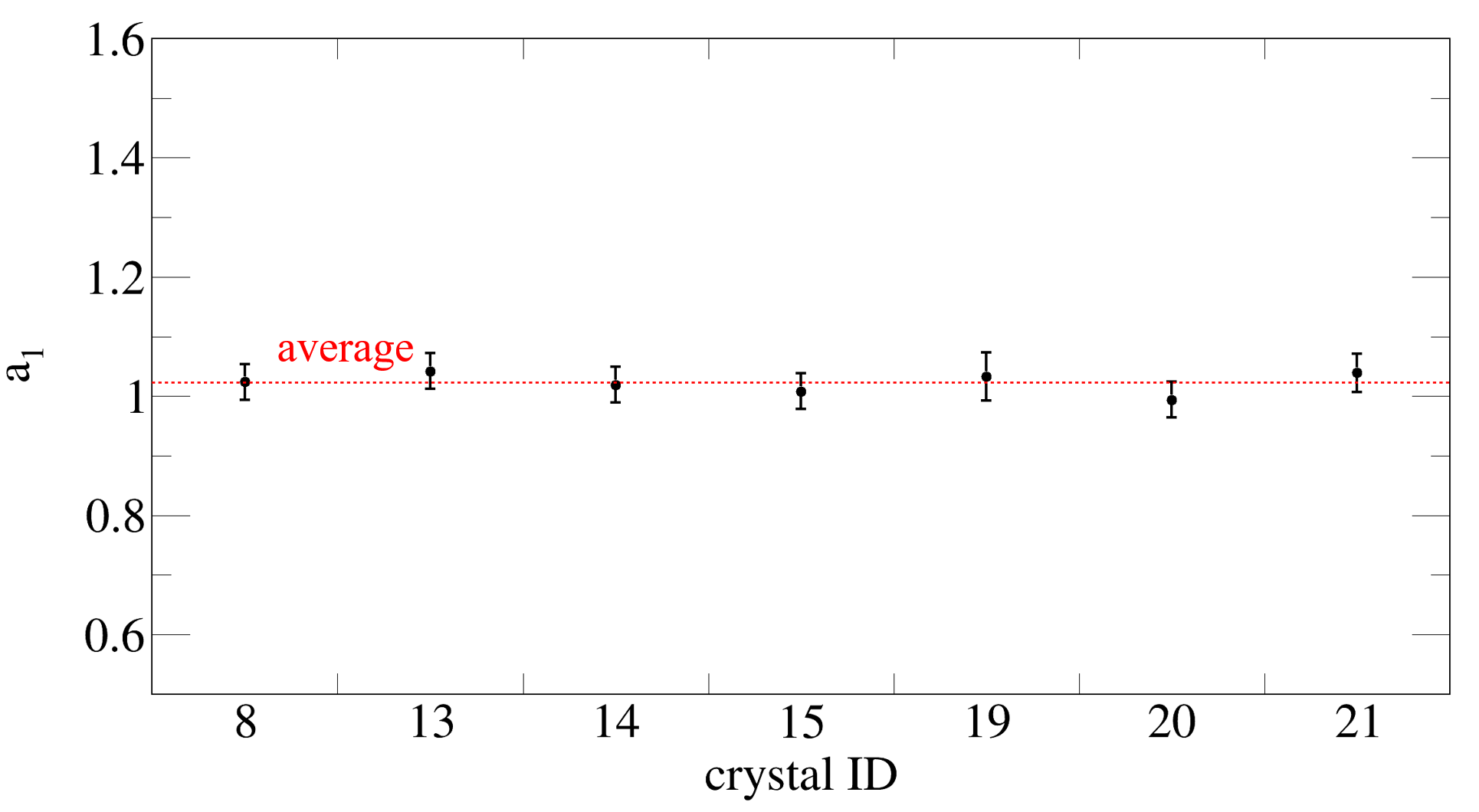}
	\caption{$a_1$ values from the modified Birks fits for seven GAGG(Ce) crystals near the center of the array. The error bars indicate the fit uncertainties. The red dashed line indicates the average $a_1$ value for the selected 7 crystals.}
	\label{figa1com}
\end{figure}

As shown in Fig.~\ref{figa1com}, the extracted $a_1$ values are all close to 1, indicating that the quenching behavior is highly consistent among these crystals from the same production batch. The red dashed line in Fig.~\ref{figa1com} indicates the average $a_1$ value for the selected 7 crystals. This consistency suggests that the fitted parameters derived in this work are representative of the response of the GAGG(Ce) crystals, rather than reflecting an isolated case of a single detector. Consequently, the obtained parameterization can be applied to similar GAGG(Ce) scintillators for response estimation and quenching correction in future charged-particle detection experiments under comparable conditions.

\subsection{Discussion}

\begin{figure}[!htb]
	\includegraphics[width=\hsize]{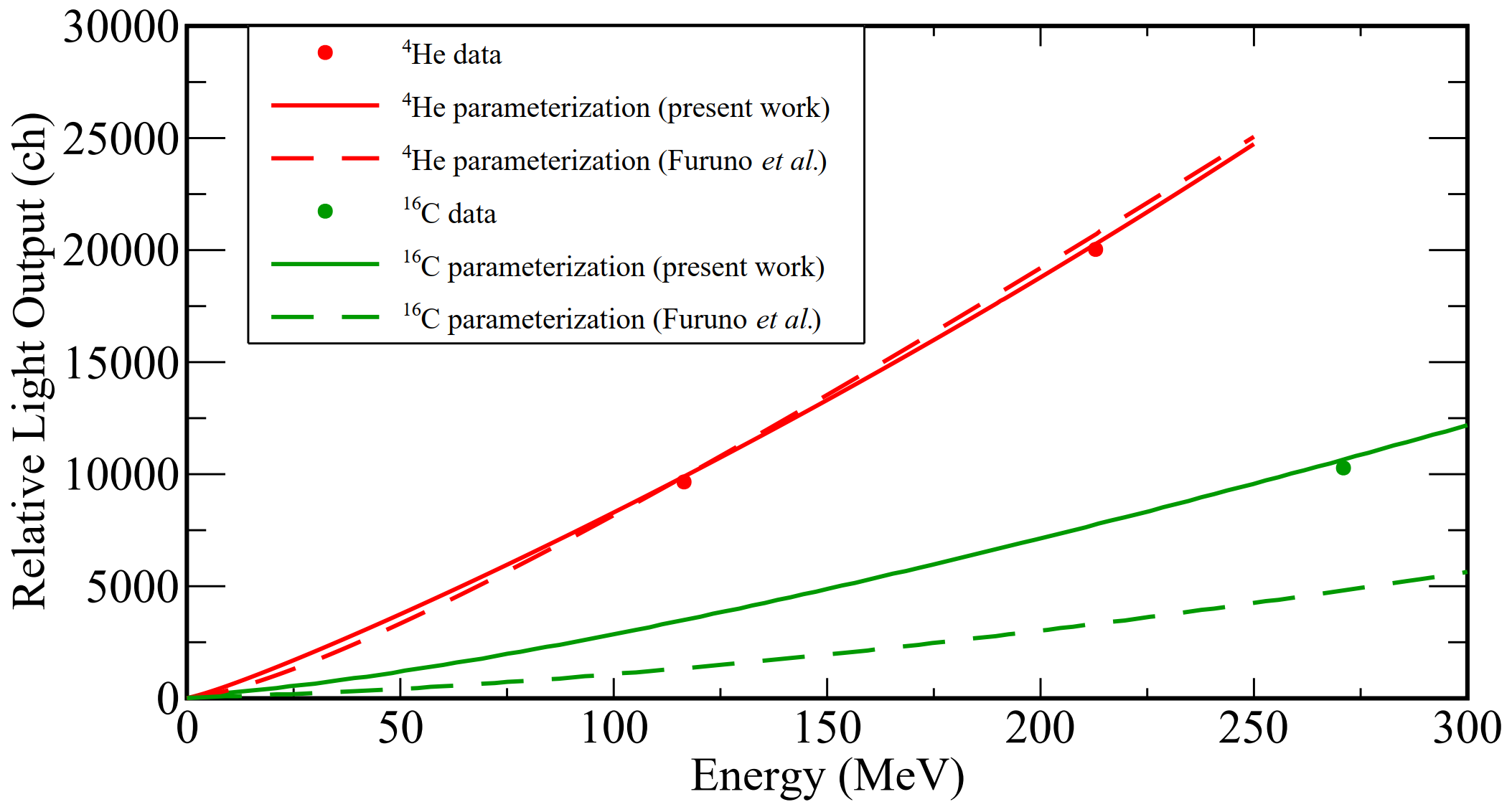}
	\caption{Comparison of the relative light output of GAGG(Ce) obtained using the parameterization from the present work (solid lines) with that reported by Furuno \textit{et~al.}~\cite{bib:18}(dashed lines) for \ce{^4He} and \ce{^{16}C}. Solid circles represent the experimental data from the present work.}
	\label{figfurunocom}
\end{figure}

Previous studies of GAGG(Ce) have largely focused on light ions, mainly protons and $\alpha$ particles, at relatively low energies, providing limited reference data for the higher deposited energies and heavier nuclei relevant to quasi-free knockout-reaction measurements with radioactive ion beams. Furuno \textit{et~al.} reported measurements for protons with energies of \mbox{\SIrange{5}{68}{\MeV}} and $\alpha$ particles with energies of \mbox{\SIrange{8}{54}{\MeV}} in~\mbox{GAGG(Ce)}~\cite{bib:18}. Fig.~\ref{figfurunocom} shows a comparison between their results and the present work. To remove the overall scale difference, both results were normalized to the light output of \ce{^4He} at a deposited energy of $E_{\mathrm{dep}} \approx \SI{116.5}{\mega\electronvolt}$. In addition, the relative light output of \ce{^{16}C} as a function of deposited energy is included as a representative example of a heavier nucleus at higher energy.

After normalization, the two results are quite close for \ce{^4He}, both showing good agreement with the data. In contrast, clear deviations are observed for heavier charged particles and at higher deposited energies. In particular, for \ce{^{16}C}, the relative light output calculated using the parameterization of Furuno \textit{et~al}. is significantly lower than the measured result, whereas the present parameterization reproduces the data more consistently over the investigated range. The parameterization reported by Furuno \textit{et~al}. was primarily derived from measurements involving low-energy protons and $\alpha$ particles. As a result, it does not adequately reproduce the response observed in the present work for heavier ions such as carbon. The discrepancy is attributed to the higher linear energy transfer and extended energy range investigated here. In contrast, the present parameterization provides a more consistent description across different particle species and energies.

\section{Energy resolution}\label{sec.V}
The light-output response functions obtained in the previous sections provide an energy calibration for the GAGG(Ce) scintillators. Based on this calibration, the energy resolution of the GAGG(Ce) crystals was evaluated using the \ce{^4He} beam.

By restricting the energy spread of the secondary \ce{^4He} beam, the particles entering the telescope were prepared with an approximately monoenergetic energy distribution, enabling a reliable determination of the energy resolution of the GAGG(Ce) scintillators. Events corresponding to particles incident nearly normal to the scintillator surface were selected using the position information provided by the DSSD, thereby minimizing variations in the incident angle and positions and thus the associated uncertainties in energy deposition.

\begin{figure}[!htb]
	\includegraphics[width=\hsize]{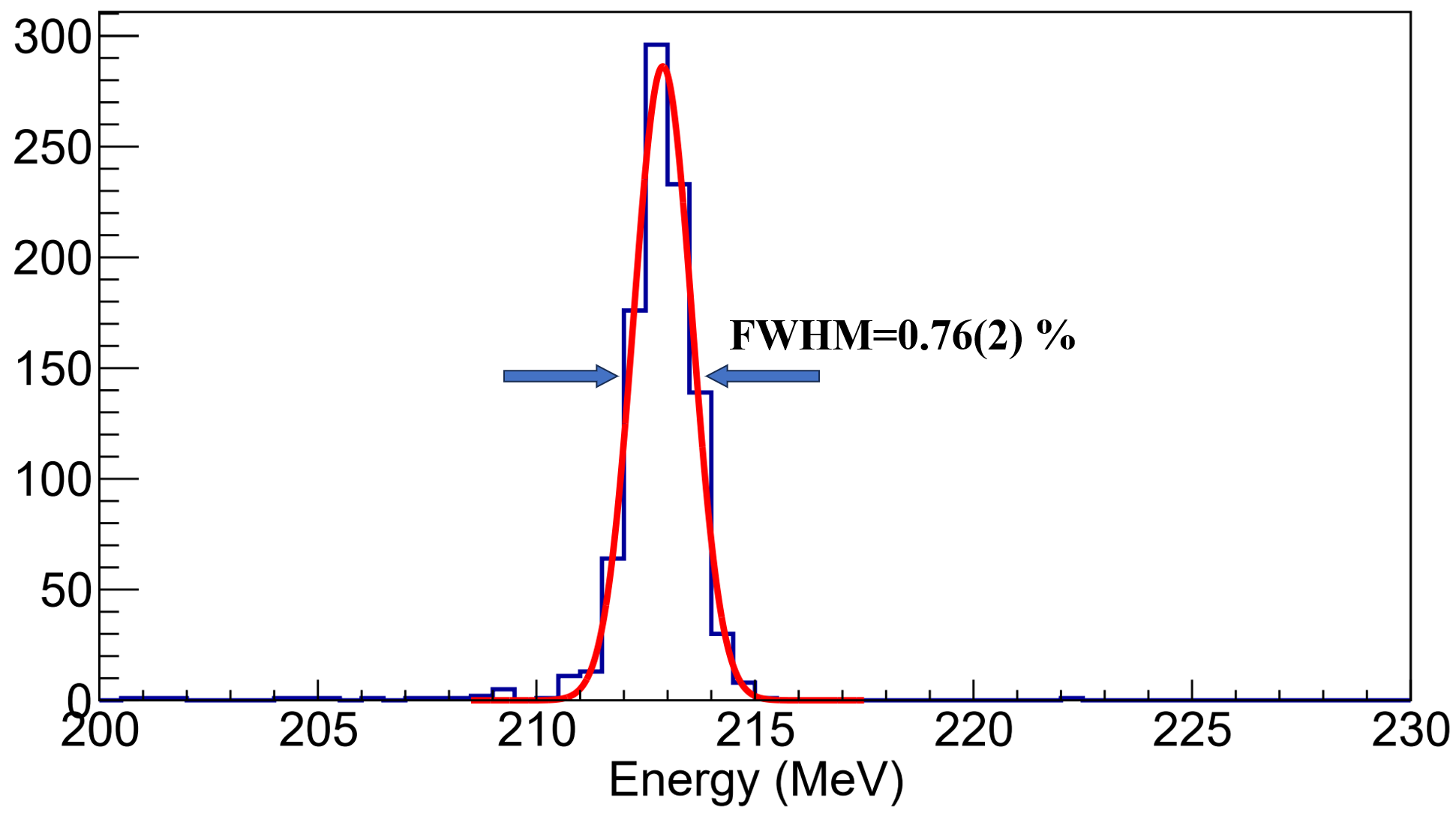}
	\caption{Energy spectrum measured with a GAGG(Ce) scintillator for a \SI{212.9}{\MeV} \ce{^4He} beam. The red line represents the Gaussian fit, giving an energy resolution of 0.76(2)\%.}
	\label{figresolution}
\end{figure}

Under these conditions, the energy resolution of the GAGG(Ce) crystals in the array was determined for the \SI{212.9}{\MeV} \ce{^4He} beam. As a representative example, the energy spectrum measured with the GAGG(Ce) crystal located at the center of the array (crystal No.~15) is shown in Fig.~\ref{figresolution}, together with the corresponding Gaussian fit. An energy resolution of 0.76(2)\% (FWHM) was obtained for this crystal, demonstrating that the GAGG(Ce) scintillators provide sufficient energy resolution for the requirements of the present and future charged-particle detection experiments.It should be noted that this value includes the energy spread of the \ce{^{4}He} beam, which is estimated to be at the level of around 0.5\%. The geometrical contribution is minimized by selecting centrally and nearly normally incident particles using the DSSD position information.

\section{Summary}
In this work, the scintillation light-output response of GAGG(Ce) scintillators to various charged particles was systematically investigated. Beams of \ce{^{4}He}, \ce{^{10}Be}, \ce{^{10}C}, \ce{^{11}C}, \ce{^{16}C}, and \ce{^{18}O} were used to cover a broad range of deposited energies and linear energy transfers. The measured light-output response was analyzed using both the standard Birks model and a modified Birks parameterization. While the standard Birks model provides a qualitative description of the response, the modified Birks model yields a significantly improved agreement with the experimental data. The resulting fit parameters of the modified Birks model were found to be consistent across multiple GAGG(Ce) crystals from the same production batch, indicating good reproducibility and enabling a transferable description of the scintillation response, particularly the quenching behavior. Based on the established calibration, the energy resolution of the GAGG(Ce) crystals was evaluated using a \ce{^4He} beam with a deposited energy of \SI{212.9}{\MeV}. An overall experimental energy resolution of 0.76(2)\% (FWHM) was obtained, including the contribution from the beam energy spread. These results provide useful experimental input for response modeling, light-output calibration, and energy reconstruction of GAGG(Ce)-based charged-particle detector systems.

\section*{Generative AI Statement}
During the preparation of this work, the authors used ChatGPT-4o for the purpose of language editing and grammar checking only. The tool was used to improve the readability and clarity of the text. After using this tool, the authors reviewed and edited the content as needed and take full responsibility for the content of the published article. No AI tool was used to generate scientific data, perform analysis, or formulate conclusions.

\begin{acknowledgments}
	We would like to thank the  HIRFL-RIBLL staff and the LENG (Low temperature target system for Exotic Nuclei reactions with Gas) collaboration for the technical support. This work was supported by the National Key R\&D Program of China (Grants No. 2023YFE0101500, No. 2022YFA1605100), the National Natural Science Foundation of China (Grant Nos. 12275006 and 12175280), the State Key Laboratory of Nuclear Physics and Technology, Peking University (Nos. NPT2024ZX01 and NPT2025ZX02), the Research Program of Heavy Ion Science and Technology Key Laboratory, Institute of Modern Physics, Chinese Academy of Sciences (Grant No.~HIST2025KS01), the CAS “Global Initiative on Common Challenges" (Grant No.~016GJHZ2023063GC), Major Science and Technology Projects in Gansu Province (Grant No. 24GD13GA005).
\end{acknowledgments}


\end{document}